\documentclass[twocolumn,showpacs,preprintnumbers,nofootinbib,floatfix]{revtex4}
\usepackage{graphicx}
\usepackage{dcolumn}
\usepackage{bm}
\usepackage{amsfonts}
\usepackage{amsthm}
\usepackage{amsmath}
\usepackage{amssymb}
\usepackage{epsfig}
\include{epsf}

\usepackage{times}

\def\beq{\begin{equation}}
\def\eeq{\end{equation}}
\def\bey{\begin{eqnarray}}
\def\eey{\end{eqnarray}}

\def\lsim{\mathrel{\raise.3ex\hbox{$<$\kern-.75em\lower1ex\hbox{$\sim$}}}}
\def\gsim{\mathrel{\raise.3ex\hbox{$>$\kern-.75em\lower1ex\hbox{$\sim$}}}}
\def\Msun{M_\odot}

\newcommand{\bea}{\begin{eqnarray}}
\newcommand{\eea}{\end{eqnarray}}
\newcommand{\be}{\begin{equation}}
\newcommand{\ee}{\end{equation}}
\newcommand{\bi}{\begin{itemize}}
\newcommand{\ei}{\end{itemize}}

\preprint{UTTG-04-11  \;\;\;\;\;\;\; TCC-006-11}

\begin{document}
\title{Constraints on a Non-thermal History from Galactic Dark Matter Spikes}
\author{Pearl Sandick}
\email{pearl@ph.utexas.edu}
\affiliation{Theory Group and Texas Cosmology Center, 
The University of Texas at Austin, TX 78712}
\author{Scott Watson}
\email{gswatson@syr.edu} 
\affiliation{Department of Physics, 
Syracuse University, Syracuse, NY  13244.}
\date{\today}
\begin{abstract}
In this paper we examine whether indirect detection constraints on dark matter associated with a non-thermal history may be significantly improved when accounting for the presence of galactic substructure in the form of dark matter spikes.  We find that significant constraints may be derived from the non-observation of an excess of diffuse gamma-rays and from the properties of bright gamma-ray point sources observed by the Fermi Gamma-Ray Space Telescope, but these constraints depend sensitively on
the details of the formation of the first stars and their subsequent black hole remnants. However, we also find that, especially if WIMPs annihilate primarily to quarks or gauge bosons, it is possible to extract meaningful and conservative bounds on the annihilation cross section.
\end{abstract}
\maketitle

\section{Introduction}
Despite the successes of precision cosmology, existing observations seem to tell us little about the history of the universe prior to Big Bang Nucleosynthesis (BBN).  This is unfortunate given the expectations from particle theory for a rich amount of phenomenology at these scales; including symmetry breaking transitions, the generation of mass in the Standard Model, the origin of the baryon asymmetry, and the existence of Cold Dark Matter (CDM).  The Large Hadron Collider (LHC) is currently probing the microphysics responsible for many of these processes, but without an exact description of the cosmological history prior to BBN our understanding will remain incomplete.

In the case of CDM, the standard assumption of a {\em thermal} history prior to BBN provides a well-motived and convincing scenario for connecting the cosmological and microscopic origin of CDM \cite{Bertone:2004pz}.  In this approach, one assumes that very early in its history the universe achieves thermal equilibrium and remains in that state until the time of BBN.  In such an approach, the amount of CDM today depends parametrically on the properties of the CDM particles (mass and cross section) and the temperature at which the particles ceased to annihilate -- so-called `freeze-out'.  It is reassuring that when comparing this estimate with precision cosmological measurements for the amount of CDM today we get a prediction for the mass and annihilation cross section near the scale of electroweak symmetry breaking.  However, in the simplest models this ``WIMP Miracle'' is spoiled by a tension with electroweak precision constraints.
Given that search strategies at LHC and other CDM detection experiments depend on assumptions about the self-annihilation cross section of these particles\footnote{
There is both a direct connection, as is the case for the indirect detection of annihilation products, as well as a more implicit connection that appears in model-specific scenarios.  As an example of the latter, in the Minimal Supersymmetric extension of the Standard Model (MSSM), requiring a thermal dark matter candidate typically leads to a WIMP that is a bino-like neutralino.  However, if one drops the thermal constraint more regions of the MSSM parameter space become viable, which can lead to different possible benchmark signatures at LHC \cite{Kane:2008gb}.} it is crucial to establish the robustness of the thermal scenario and the associated cosmological constraint, i.e. $\langle \sigma v \rangle_{th} \approx 10^{-26}$ cm$^3$s$^{-1}$, as well as identifying any other viable alternatives for their production \cite{Kane:2008gb}.

One possible alternative scenario is that of a {\em non-thermal} history.  This scenario occurs if massive particle decays or phase transitions lead to a significant entropy and particle production prior to BBN.  If such transitions occur after the thermal freeze-out of CDM, predictions for the microscopic properties of the {\em total} amount of CDM may differ significantly from the usual thermal scenario \cite{Watson:2009hw}.   Such scenarios have deservingly received much skepticism over the years, particularly because the non-thermal production of CDM must occur in a very narrow window -- after CDM thermal freeze-out but prior to the onset of BBN -- naively introducing a new and unmotivated scale of physics into the problem.  However, in the particular case of Anomaly Mediated Supersymmetry (SUSY) breaking one finds that this scale is set by the scale of SUSY breaking and non-thermal CDM is a natural prediction for this class of models \cite{Moroi:1999zb}.  Building on this intuition, more recently it has been suggested that this may be a general expectation of a larger class of gravity-mediated SUSY models when one accounts for theoretical self-consistency in the ultraviolet \cite{Watson:2009hw,nontherm}. In addition to this theoretical motivation, non-thermal models make definite and testable predictions which are currently being scrutinized at colliders, as well as by ground- and space-based CDM searches.  One such prediction is the enhancement of the self-annihilation cross section by as much as three orders of magnitude compared with that of the standard thermal scenario, while still yielding the correct amount of CDM cosmologically \cite{Watson:2009hw}.  

 In this paper we focus on whether indirect detection constraints on non-thermal CDM can be significantly improved when accounting for galactic substructure resulting from dark matter spikes.  Specifically, we follow \cite{Sandick:2010yd} in their analysis of the gamma-ray constraints on dark matter annihilation in the dark matter spikes in our Galactic halo. Dark matter spikes arise due to the contraction of a dark matter minihalo when a baryonic object (e.g.~a star) forms at its center, as was the case with the first generation of stars to form in our universe.  Indeed, as a result of the increased dark matter density in the spike, the very first stars are thought to have undergone a phase during which they were supported by dark matter annihilations, dubbed the Dark Star phase~\cite{darkstars}. The affect of a boosted annihilation cross section on the evolution of Dark Stars was examined in~\cite{Ilie:2010vg}, where they found that the Dark Star phase is shortened by an enhanced dark matter annihilation cross section, though the existence of the phase is robust.

Current constraints from indirect detection already put strong bounds on the allowable cross sections for non-thermal models \cite{nonthermdet,Kane:2009if}.  Additionally, if the PAMELA\footnote{Payload for Antimatter Matter Exploration and Light-Nuclei Astrophysics} excess is in-fact a signature of dark matter annihilations, the data suggests that dark matter annihilates preferentially to leptonic final states \cite{PAM}.  It is possible to construct such models \cite{leptophilic}, however it is noteworthy that predicted fluxes of charged particles can suffer from large uncertainties associated with astrophysical backgrounds.  Indeed, it was demonstrated in \cite{Kane:2009if} (see also \cite{Grajek:2010bz,Perelstein:2010gt}) that in the case of anti-protons, astrophysical backgrounds can be significantly lower than previously expected while still being consistent with the Boron to Carbon ratio.  

Here we examine the potential of the gamma-ray data from the Fermi Gamma-Ray Space Telescope (FGST) to constrain models of non-thermal dark matter in local spikes. Uncertainties in the astrophysical backgrounds play an inconsequential role in the following analysis, however, we find that constraints on dark matter annihilation can be ambiguous in the absence of a reliable star formation history. We emphasize that an important and difficult challenge for this program is the establishment of constraints on the typical mass and formation era of the first generation of stars.  For example, upcoming observations with the James Webb Space Telescope (JWST) may provide some hints about the formation of the first stars~\cite{jwstPopIII}, and it's even possible that JWST will observe a Dark Star~\cite{jwstDS}. As we demonstrate, if a star formation history is established, the constraints on dark matter annihilation in local spikes may be very significant.

In the next section we briefly review the mechanism by which dark matter spikes form in the early universe and the method used to extract the local distribution of surviving spikes in our Galactic halo. 
In section~\ref{sec:constraints} we demonstrate how the presence of this substructure can lead to a general enhancement of the gamma-ray constraints on non-thermal dark matter model building.  We briefly conclude in the section that follows. 

\section{Galactic Substructure from CDM Spikes}
\label{sec:spikes}
One might expect that the distribution of CDM within our galaxy can be strongly influenced by the formation and evolution of objects such as black holes.  In particular, 
as the gravitational potential becomes dominated by a compact baryonic object, the CDM distribution near this object will be affected.
Gondolo and Silk have examined this possibility for the supermassive back hole at the center of our galaxy (around $10^6$ M$_\odot$) around which one might expect a large enhancement in the CDM density \cite{Gondolo:1999ef}.  However, further investigations revealed that such extreme inhomogeneities are most likely negligible today due to a number of effects, including major merger events, off-center formation of the seed black hole, gravitational scattering off stars, and CDM annihilations \cite{Ullio:2001fb,Merritt:2002vj,Bertone:2005hw,Bertone:2005xv}.
Zhao and Silk then proposed \cite{Zhao:2005zr} that these wash-out effects may not be present for small over-densities, or {\em spikes}, resulting from Intermediate Mass Black Holes (IMBHs), which are the expected remnants of the earliest stars to form, known as Pop-III stars\footnote{In this paper we will carelessly refer to Population III.1 as Pop-III.}.  

Bertone, Zentner, and Silk (BZS) examined this possibility in more detail in \cite{Bertone:2005xz} (for a review see \cite{Bertone:2009kj}) using an analytic model of halo evolution and performing $200$ statistical realizations for the growth of a Milky Way-sized halo.  The population of IMBHs was generated by identifying  $3\sigma$ over-densities in the smoothed primordial density field at a redshift of \mbox{$z=18$} and replacing each of those peaks with a $100$ M$_\odot$ black hole.  
Tracking the growth and mergers of the structures until today, they find an expected number of IMBHs in our galaxy to be $N_{{bh}}=1027 \pm 84$.  The uncertainty in this number reflects unknowns in the model parameters, such as the redshift at which small-scale fragmentation of baryonic disks becomes important and black hole seeds cease to form.  BZS accounted for this uncertainty by varying the redshift at which the seeds are initially evolved.

A different approach was taken by one of us (PS) in collaboration with J. Diemand, K. Freese, and D. Spolyar in \cite{Sandick:2010yd} (see also \cite{Sandick:2010qu}), hereafter referred to as SDFS. Their analysis uses the Via Lactea-II cosmological N-body simulation \cite{Diemand:2008in} to estimate the number and mass distribution of CDM minihalos as a function of redshift.  Minihalos suitable for the formation of Pop-III stars are identified at high redshift, and the distribution evolved until today. Assuming each minihalo hosted a Pop-III star, these minihalos exist today as spikes, each surrounding an IMBH Pop-III remnant.  The distribution of these IMBHs and surrounding spikes depends on the duration of Pop-III star formation, though the exact redshift at which Pop-III star formation ceases remains uncertain. Additionally, the density profile of each individual spike, and therefore the expected dark matter annihilation rate, depends on the typical size of the remnant black holes.
Here we build on the analysis of SDFS, whose methodology we now review, referring to the original papers for more detail~\cite{Sandick:2010yd,Sandick:2010qu}.

Via Lactea-II \cite{Diemand:2008in} is the first cosmological N-body simulation of a Milky-Way sized dark matter halo capable of resolving the $\sim 10^6$ M$_\odot$ minihalos in which the first stars formed.
Star formation depends on the ability of the baryonic clouds to efficiently cool as they collapse. This cooling proceeds primarily through excitations of molecular hydrogen, the abundance of which depends on the temperature and therefore redshift of formation.  Using this fact, Trenti and Stiavelli found a minimal mass for minihalos in which Pop-III stars could have formed of \cite{Trenti:2009cj} 
\be
M_{{min}}^{{halo}}=1.54 \times 10^5 \; \Msun \left(  \frac{1+z}{31} \right)^{-2.074}.
\ee
The maximum mass of halos that formed Pop-III stars is less important since the hierarchical nature of structure formation favors small mass minihalos, but for completeness SDFS took a maximum mass of $10^7$ M$_\odot$.  
Given the uncertainty in the redshift at which Pop-III star formation gave way to the formation of less-massive subsequent generations of stars (which are not expected to result in the spikes we examine here), SDFS considered three possible termination redshifts $z_f=11$, $15$, and $23$.  For brevity, here we consider only $z_f=15$.

Assuming each Pop-III star ended its life by collapsing to a black hole, and given a Pop-III termination redshift and the viable minihalo mass range above, 
the current number density of black holes surrounded by spikes, $N_{bh}$, is related to the total possible number of viable minihalos, $N_{halos}$, by
\be \label{numberofbh}
N_{bh}= f_0 \left( 1- f_{merged} \right) N_{halos},
\ee
where $f_0$ is the fraction of halos that are expected to host Pop-III stars, and $f_{merged}$ is the fraction of CDM spikes that are destroyed by black hole mergers.  SDFS argued that mergers are most important for the highest mass black holes and for $f_0 \approx 1$, in which case they would reduce the number of spikes by at most a factor of two.  For lighter black holes and/or smaller $f_0$, it was argued that this effect is negligible and $N_{bh} \approx f_0 N_{halos}$. Here we will fix the fraction of black holes to form and survive, $f_s=f_0(1-f_{merged})$ and consider two possible values; $f_s=0.1$ and the maximal case $f_s=1$.

If the growth rate of a baryonic object at the center of a minihalo is slow with respect to the time it takes CDM particles to cross the central region, the contraction of particle orbits and the formation of CDM spikes may be modeled by adiabatic contraction.  SDFS used the Blumenthal {\it et al.} prescription for adiabatic contraction \cite{Blumenthal:1985qy}, which predicts a roughly power-law density profile for the spikes and is independent of the CDM particle mass. However, given the enhanced density of CDM, some particle self-annihilations will take place.  This depends on the lifetime of the central mass core ($t_{bh}$) and leads to an upper limit on the CDM density 
\be
\rho_{max}=\frac{m_{\chi}}{\langle \sigma v \rangle t_{bh}},
\ee
where $m_{\chi}$ and $\langle \sigma v \rangle$ are the CDM mass and averaged self-annihilation cross section times velocity, respectively. 

In summary, SDFS find that the formation of Pop-III stars leads to a significant number of CDM spikes in our own galaxy today, as first anticipated by Zhao and Silk.  In the next section, we consider the feasibility of using the existence of these spikes to sharpen constraints on the properties of CDM through (non-)observation of their annihilation products by FGST. 

\section{Non-thermal CDM Constraints from Spikes}
\label{sec:constraints}
For a Majorana CDM particle with mass $m_\chi$ and average annihilation cross section times velocity $\langle \sigma v \rangle$, the rate of self annihilations inside a spike is 
\beq  
\Gamma = \frac{\langle \sigma v \rangle}{2 m_\chi^2}\int_{r_{min}}^{r_{max}} dr \, 4 \pi r^2 \, \rho_{spike}^2(r),
\label{eq:rate}
\eeq
where $r_{min}$ and $r_{max}$ are the inner and outer radii of the CDM spike in which annihilations occur with the former being of order the Schwarzschild radius of the black hole, and $\rho_{spike}(r)$ is the CDM density profile of the spike.

We consider several WIMP candidates defined by their masses and annihilation channels.  Calculations are performed for WIMP masses of 100, 200, 500, 1000, and 2000 GeV and Standard Model final states $b \bar{b}$, $W^+W^-$, $\tau^+ \tau^-$, and $\mu^+ \mu^-$. 
The resulting spectrum of photons $dN_f/dE$ from annihilation to final state $f$ is computed with PYTHIA~\cite{pythia}.  
For $\chi \chi \rightarrow \mu^+\mu^-$, the photon spectrum comes from final state radiation and is given by~\cite{fsr}

\beq
\frac{dN_{\mu^+\mu^-}}{dx} = \left(\frac{x^2-2x+2}{x\pi / \alpha}\right) 
\left[ \ln \left(\frac{s(1-x)}{m_\mu^2}\right)-1\right],
\eeq
where $x\equiv E_\gamma/m_\chi$, the center-of-mass energy squared is $s=4m_\chi^2$, and $\alpha \approx 1/137$ is the fine structure constant. We note that WIMP candidates typically annihilate to a variety of final states with the rate of annihilations in a CDM spike expressed as
\beq
\Gamma_f= B_f \Gamma,
\label{eq:rate2}
\eeq
where $B_f$ is the branching ratio to the final state $f$.
The intrinsic photon luminosity from CDM annihilations in any CDM spike is then
\beq
{\cal L} = \int dE \,\sum_f \frac{dN_f}{dE} \,\Gamma_f.
\eeq 
Given the luminosity resulting from dark matter annihilations in spikes, we can now proceed to establish constraints on the CDM self annihilation cross section using both point source and diffuse flux data from FGST. 

\subsection{Point Source Constraints}
We first consider establishing constraints on the WIMP self annihilation cross section by requiring that annihilations in the nearest spike do not lead to a point source flux that exceeds that from the brightest recorded FGST point source. Point source constraints rely heavily on the estimate of the distance to the nearest spike, determined by integrating the probability density of finding a spike in the neighborhood of our Solar System. Despite the fact that the brightest FGST point source is associated with the Vela pulsar~\cite{fgstFSC}, in this analysis we simply require that the gamma-ray flux from the brightest spike not exceed the gamma-ray flux from Vela, resulting in the somewhat bizarre requirement that the brightest spike must be located along our line-of-sight to Vela.  As this possibility is not excluded, we reserve further discussion of this issue until the end of the section. 

\begin{figure}[h!]
\begin{center}
\mbox{\epsfig{file=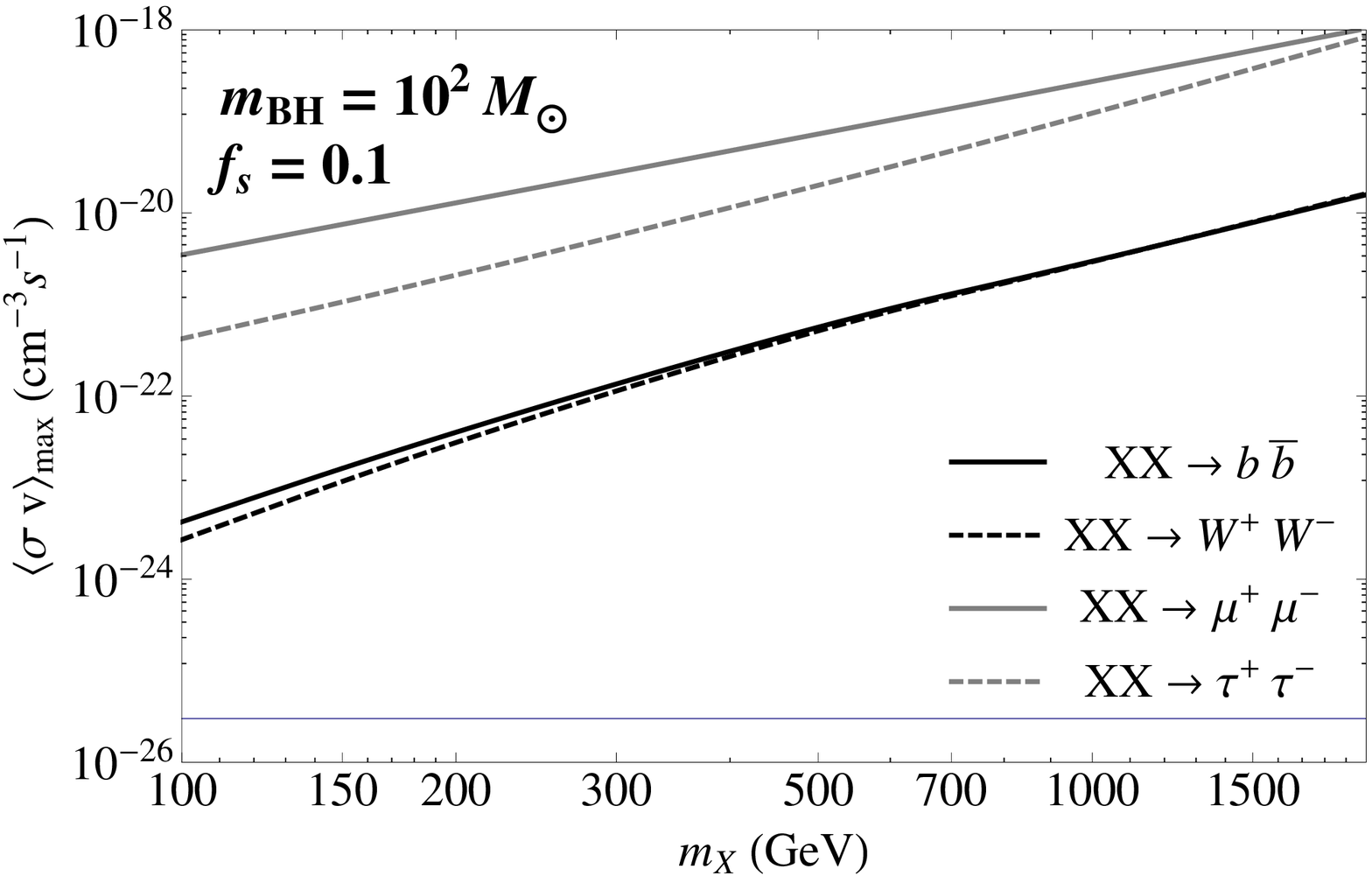,width=.48\textwidth}}\vspace{5mm}
\mbox{\epsfig{file=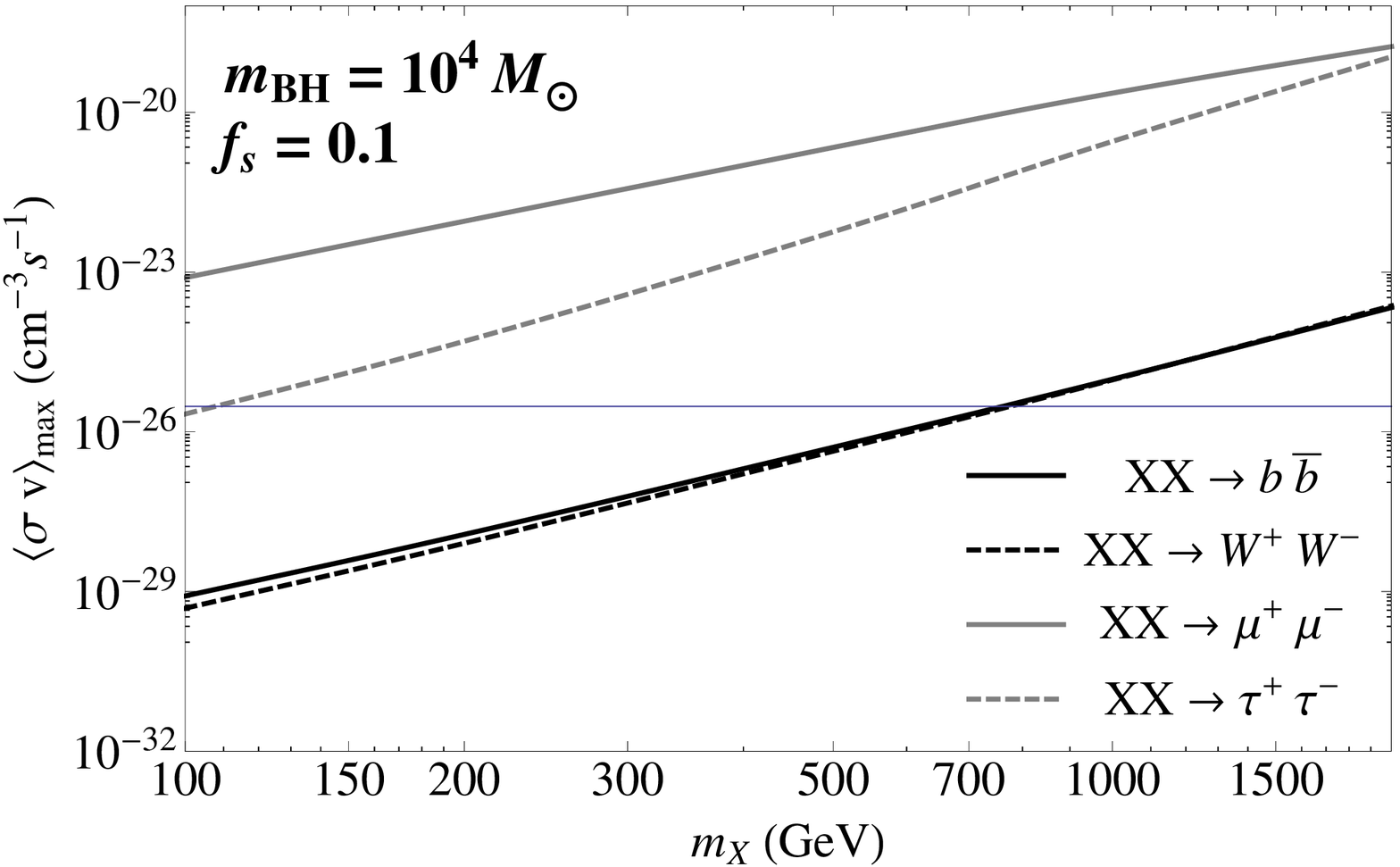,width=.48\textwidth}}
\end{center}
\caption{\it Upper limit on the annihilation cross section as a function of the mass of the dark matter particle for a typical black hole mass of $10^2$ M$_\odot$ (top panel) and $10^4$ M$_\odot$ (bottom panel) for four choices of final state particles; $b \bar{b}$ (solid black curves), $W^+W^-$ (dashed black curves), $\mu^+\mu^-$ (solid grey curves), and  $\tau^+\tau^-$ (dashed grey curves). Here we assume $f_s=0.1$. The horizontal line in each panel indicates $\langle \sigma v \rangle_{th}=3 \times 10^{-26}$ cm$^3$s$^{-1}$.
\label{fig:PSpt1}}
\end{figure}

\begin{figure}[h!]
\begin{center}
\mbox{\epsfig{file=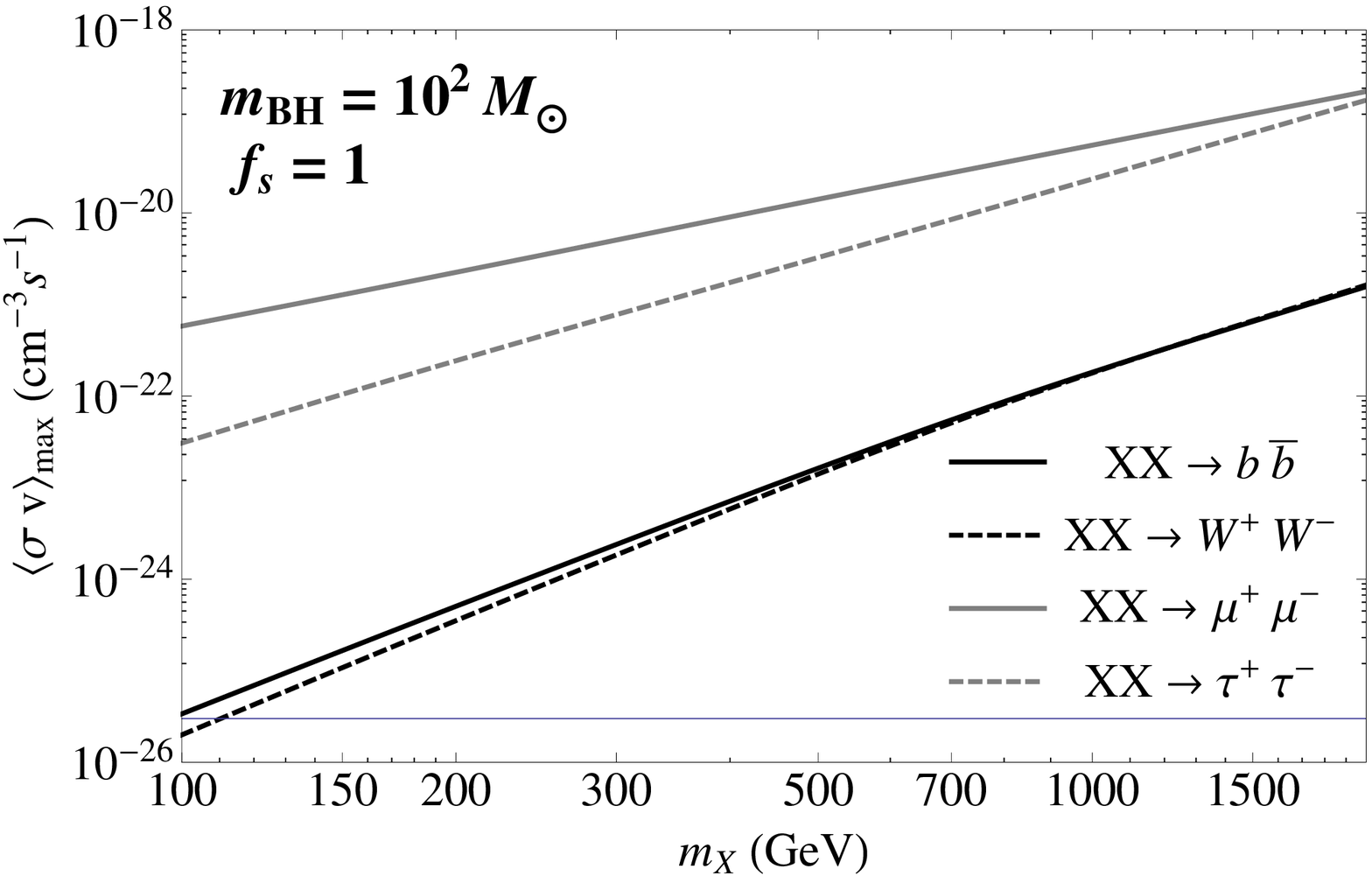,width=.48\textwidth}}\vspace{5mm}
\mbox{\epsfig{file=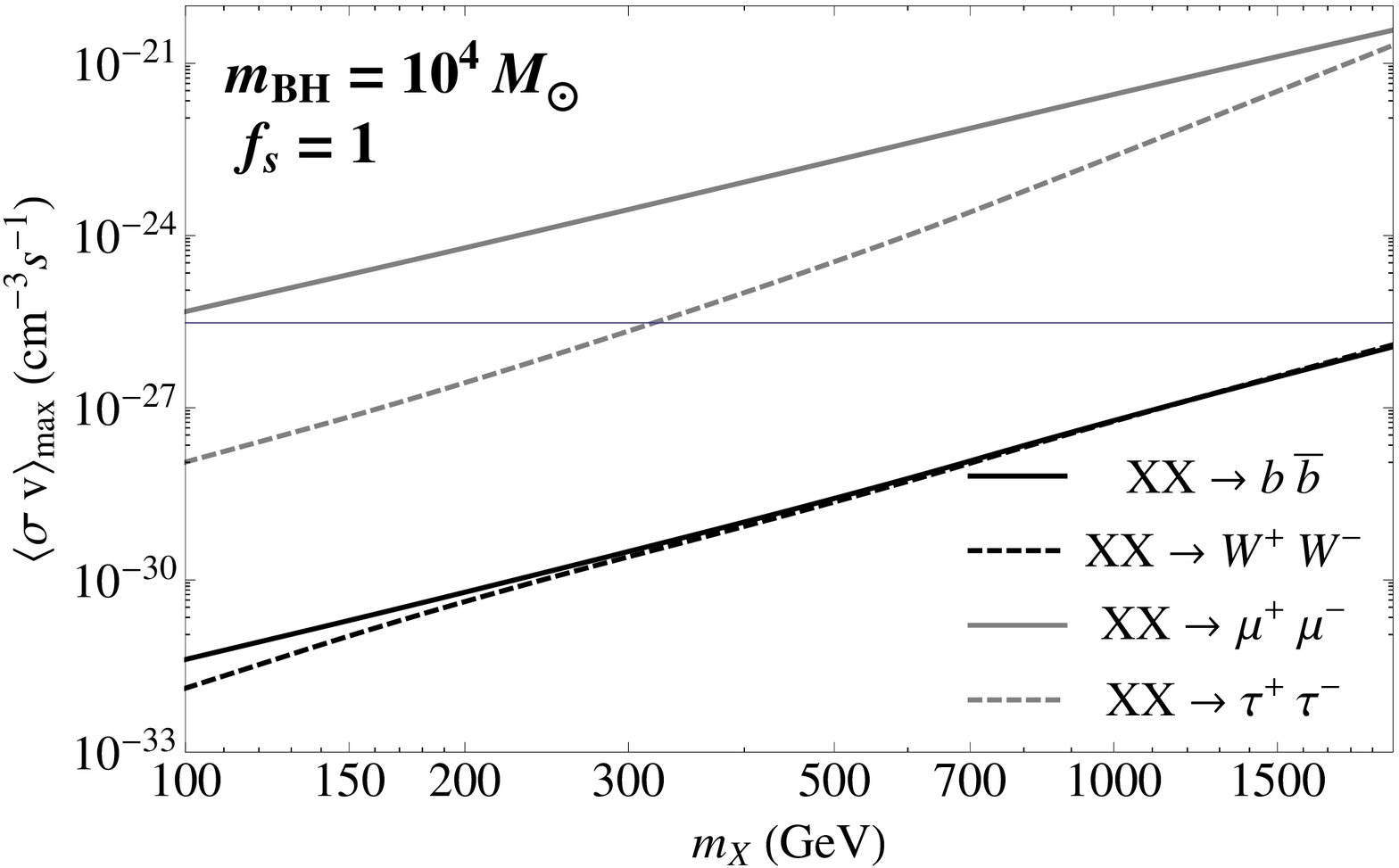,width=.48\textwidth}}
\end{center}
\caption{\it Upper limit on the annihilation cross section as a function of the mass of the dark matter particle for a typical black hole mass of $10^2$ M$_\odot$ (top panel) and $10^4$ M$_\odot$ (bottom panel) for four choices of final state particles; $b \bar{b}$ (solid black curves), $W^+W^-$ (dashed black curves), $\mu^+\mu^-$ (solid grey curves), and $\tau^+\tau^-$ (dashed grey curves). Here we assume $f_s=1$. The horizontal line in each panel indicates $\langle \sigma v \rangle_{th}=3 \times 10^{-26}$ cm$^3$s$^{-1}$.
\label{fig:PS1}}
\end{figure}

As discussed in section~\ref{sec:spikes}, given the uncertainties in spike formation we will consider both small mass ($m_{BH}=100$ M$_\odot$) and large mass ($m_{BH}=10^4$ M$_\odot$) black holes and we will consider two values for the fraction of black holes to form and survive; $f_s=0.1$ and $f_s=1$.  Our results for the point source analysis are given in Figures~\ref{fig:PSpt1} and~\ref{fig:PS1}.  In Figure~\ref{fig:PSpt1} 
we fix $f_s=0.1$ and present the upper limit on the average WIMP annihilation cross section times velocity as a function of the mass of the dark matter particle for a typical black hole mass of $100$ M$_\odot$ (top panel) and $10^4$ M$_\odot$ (bottom panel) for four choices of final state particles; $b \bar{b}$ (solid black curves), $W^+W^-$ (dashed black curves), $\mu^+\mu^-$ (solid grey curves), and $\tau^+\tau^-$ (dashed grey curves). In each panel we present the cosmologically determined thermal WIMP cross section $\langle \sigma v \rangle_{th}=3 \times 10^{-26}$ cm$^3$s$^{-1}$ for comparison.  Similarly, in Figure~\ref{fig:PS1}, we show the upper limit on $\langle \sigma v \rangle$ as a function of WIMP mass for $f_s = 1$. 

It is clear that the constraints are quite sensitive to both the dark matter annihilation mode and the typical black hole mass.  Constraints are also sensitive to the value of $f_s$, an as-yet unknown parameter.
The least constrained case considered here is shown in the top panel of Figure~\ref{fig:PSpt1}, with $m_{bh}=100$ M$_\odot$ and $f_s=0.1$.
Comparison of these limits with the left panel of Figure~5 in \cite{Abdo:2010dk} reveals that the constraints are comparable\footnote{Differences in the slope of the constraint in the $(m_\chi,\langle \sigma v \rangle)$ plane are due in part to the fact that the spike profiles (specifically $\rho_{max}$) are affected by the WIMP mass.  As WIMP mass increases, not only does the luminosity of a particular spike decrease due to the decreased number density of WIMPs, but the luminosity also decreases because the spike has essentially lost more of its core.}.
However, for the 
heaviest black holes considered here and $f_s=1$, as shown
in the lower panel of Figure~\ref{fig:PS1}, we see that non-thermal models with light WIMPs are essentially completely ruled out, with the possible exception of the case where $\chi \chi \rightarrow \mu^+\mu^-$.  We expect that the true effect of the spikes will lie somewhere between these two extreme cases.  However, we see from the upper panel of Figure \ref{fig:PSpt1} that even in this case
meaningful constraints may be achieved for WIMPs with final states of $b \bar{b}$ and $W^+W^-$.  The latter places an important constraint on the non-thermal wino-like LSP scenario discussed in \cite{Kane:2009if,Grajek:2008jb}, which used the non-thermal enhancement of the cross section to address the PAMELA data.

Finally, it is obvious that the constraints presented here would improve had we chosen to use the flux limit from the brightest unassociated FGST point source, rather than the flux limit from Vela. For a spike located at some distance $D$ from our solar system, \mbox{$\Phi \propto \langle \sigma v \rangle / D^2$}, so it is possible to translate the constraints on $\langle \sigma v \rangle$ from Vela to constraints from the brightest unassociated source, which has an integrated luminosity $\sim 1/22$ that of Vela. Therefore the limits from the brightest unassociated source, for the choices of $f_s$ in Figures~\ref{fig:PSpt1} and~\ref{fig:PS1} would simply be shifted to lower $\langle \sigma v \rangle$ by a factor of $\sim1/22$, representing a notable improvement in the ability to constrain non-thermal cross sections.  Alternatively, one could imagine that $f_s$ is in fact less than the minimal value of $0.1$ that we have chosen to examine here. If we derive our limits according to the flux from the brightest unassociated FGST point source, the constraints in Figures~\ref{fig:PSpt1} and~\ref{fig:PS1} would apply to $f_s\approx0.001$ rather than 0.1 (upper panels), and $f_s\approx0.01$ rather than 1 (lower panels).  If it is true that there are no bright spikes located along our line of sight to any of the brightest associated FGST point sources, then even if $f_s$ is quite small there are very significant limits on $\langle \sigma v \rangle$ from non-observation of bright nearby CDM spikes. 

With these examples of how point source flux from WIMP annihilations in spikes can be used to place constraints on non-thermal dark matter, we now turn to the case of the diffuse flux.

\subsection{Diffuse Flux Constraints}
\begin{figure}[h!]
\begin{center}
\mbox{\epsfig{file=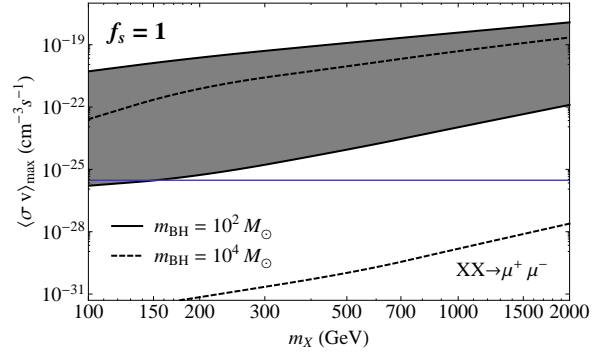,width=.48\textwidth}}
\end{center}
\caption{\it Cross sections that may be excluded by the FGST measurement of the diffuse gamma-ray flux for $f_s =1$ and annihilations to $\mu^+ \mu^-$. If the typical size for a black hole is $100$ M$_\odot$, the shaded region between the solid contours is excluded.  If the typical size for a black hole is $10^4$ M$_\odot$, the region between the dashed contours is excluded. The horizontal line indicates the thermal dark matter cross section $\langle \sigma v \rangle_{th}=3 \times 10^{-26}$ cm$^3$s$^{-1}$.
\label{fig:Diff1mu}}
\end{figure}
The CDM annihilation cross section may also be constrained by requiring that the diffuse flux from dark matter annihilations in the spikes not exceed the FGST-measured diffuse gamma-ray flux by more than $3\sigma$ in any of the nine energy bins of Ref.~\cite{fgstEGB}.
This provides a quite robust constraint given our adoption of the conservative assumptions for the number of spikes that contribute to the diffuse flux as outlined in Ref.~\cite{Sandick:2010qu}.  
In Figure~\ref{fig:Diff1mu} we present the values of $\langle \sigma v \rangle$ that may be excluded by the diffuse gamma-ray flux as measured by FGST for the channel $XX \rightarrow \mu^+ \mu^-$ for $f_s=1$. The shaded region represents the cross sections that are excluded if the typical black hole mass is $100$ M$_\odot$, while the region between the dashed contours is excluded if the typical black hole mass is $10^4$ M$_\odot$. As expected, cross sections that are below the accessible range result in too low a photon flux to provide a meaningful constraint.  However, cross sections above the accessible range result in extremely bright spikes, such that many or most of the spikes in our Galactic halo would be visible as point sources, and therefore very few would contribute to the diffuse flux. The horizontal line represents the standard cross section for thermal dark matter, $\langle \sigma v \rangle_{th}=3 \times 10^{-26}$ cm$^3$s$^{-1}$.  We see that even for the case of small black hole mass that light non-thermal WIMPs with the $XX \rightarrow \mu^+ \mu^-$ channel may lead to significant constraints.  We remind the reader, however, that the largest diffuse flux is expected in models where the luminosity of an individual spike is very low (thus many/most spikes contribute to the diffuse flux). Therefore, of the cases considered here, $XX\rightarrow \mu^+\mu^-$ with $f_s=1$ results in the strongest diffuse constraints. For $f_s<1$, the diffuse constraint weakens.  As was concluded in~\cite{Sandick:2010yd}, there are few cases where the diffuse constraint is stronger than that from point source brightness.

\section{Conclusions}
We have examined whether significant constraints on non-thermal dark matter can be derived by accounting for the presence of dark matter spikes in our Galactic halo. We find that, despite the uncertainties in the formation of dark matter spikes and the associated black holes, meaningful constraints may be expected for $f_s \gtrsim 0.1$, and even for $f_s$ as small as $10^{-3}$ if one is willing to accept that there are no spikes hiding along our line of sight to the brightest associated gamma-ray point sources. Existing constraints on non-thermal dark matter annihilation cross sections may be improved by the non-observation of a gamma-ray flux from spikes in our Galactic halo, especially if the typical mass of a black hole at the center of a spike is rather large ($\sim10^4$ M$_\odot$). 
We have also demonstrated that constraints can be established based on the contribution of faint CDM spikes to the diffuse gamma-ray flux for the example case of $\chi \chi \rightarrow \mu^+ \mu^-$.  For this particular final state, and if each minihalo capable of forming a Pop-III star did form one, we find that non-thermal WIMPs are restricted to be quite massive, even for the lighter 100 M$_\odot$ central black holes. 

Despite the many uncertainties in the star formation history, these results are promising and merit further investigation into the importance of dark matter spikes in indirect detection of non-thermal dark matter.  However, an important and difficult challenge for this program is to better establish the typical mass and formation era of the first generation of stars. As we have shown, if a star formation history is better established, accounting for dark matter annihilation in local spikes may significantly improve existing constraints on non-thermal dark matter.

\section*{Acknowledgements}
We would like to thank J\"{u}rg Diemand for his collaboration in extracting the required distributions from Via Lactea II.  We also thank K.~Freese and M.~Perelstein for useful conversations.
P.S. is supported by the NSF under Grant Numbers PHY-0969020 and PHY-0455649.  S.W. would like to thank Cornell University for hospitality.

\end{document}